\documentstyle{mn}
\input{psfig.sty}
\newcommand{\etal}{{et al.~}}

\newcommand{\kmsmpc}{\>{\rm km}\,{\rm s}^{-1}\,{\rm Mpc}^{-1}}

\newcommand{\Lsun}{\>{\rm L_{\odot}}}

\newcommand{\beq}{\begin{equation}}
\newcommand{\eeq}{\end{equation}}

\newcommand{\mpch}{\>h^{-1}{\rm {Mpc}}}

\newcommand{\apj}{ApJ}
\newcommand{\apjs}{ApJS}
\newcommand{\aj}{AJ}
\newcommand{\mnras}{MNRAS}

\newcommand{\araa}{ARA\&A}

\newdimen\hssize
\newdimen\hdsize 
\begin{document}
            

\title[The Two-Point Correlation of Galaxy Groups]
      {The Two-Point Correlation of Galaxy Groups:
        Probing the Clustering of Dark Matter Haloes}        
\author[Yang, Mo, van den Bosch \& Jing]
       {Xiaohu Yang$^{1}$, H.J. Mo$^{1}$, 
        Frank C. van den Bosch$^{2}$, Y.P. Jing$^{3}$
        \thanks{E-mail: xhyang@astro.umass.edu}\\
      $^1$ Department of Astronomy, University of Massachusetts,
           Amherst MA 01003-9305, USA\\
      $^2$ Department of Physics, Swiss Federal Institute of
           Technology, ETH H\"onggerberg, CH-8093, Zurich,
           Switzerland\\ 
      $^3$Shanghai Astronomical Observatory; the Partner Group of MPA,
           Nandan Road 80, Shanghai 200030, China}


\date{}


\maketitle

\label{firstpage}


\begin{abstract}
  We analyze the 2-point  correlation function (2PCF) of galaxy groups
  identified from  the 2-degree Field Galaxy Redshift  Survey with the
  halo based  group finder recently  developed by Yang \etal  (2004b). 
  With  this group catalogue  we are  able to  estimate the  2PCFs for
  systems ranging from isolated galaxies to rich clusters of galaxies.
  The real-space correlation length  obtained for these systems ranges
  from $\sim  4\mpch$ to  $\sim 15\mpch$, respectively.   The observed
  correlation  amplitude  (and the  corresponding  bias  factor) as  a
  function of group abundance is well reproduced by associating galaxy
  groups with dark  matter haloes in the standard  $\Lambda$CDM model. 
  Redshift  distortions  are clearly  detected  in the  redshift-space
  correlation  function, the degree  of which  is consistent  with the
  assumption of  gravitational clustering and halo bias  in the cosmic
  density field.  In agreement with  previous studies we find a strong
  increase  of  the  correlation  length  with  the  mean  inter-group
  separation.  Although well  determined observationally, we show that
  current theoretical predictions are not yet accurate enough to allow
  for stringent  constraints on cosmological  parameters.  Finally, we
  use  our results  to explore  the power-law  nature of  the  2PCF of
  galaxies.   We  split  the   2PCF  in  1-group  and  2-group  terms,
  equivalent to the 1-halo and 2-halo terms in halo occupation models,
  and show  that the power-law form  of the 2PCF is  broken, when only
  including galaxies in the more massive systems.
\end{abstract}




\section{Introduction}

In the standard cold dark  matter (CDM) cosmogony galaxies are assumed
to  form  in  virialized   dark  matter  haloes.   Theoretically,  the
properties of the halo population  can be studied in great detail with
the  use  of high-resolution  $N$-body  simulations and  sophisticated
analytical models.   Observationally, however, dark  matter haloes can
only  be  detected  indirectly;  either through  their  gravitational
lensing of background  sources, or by using galaxies  and/or X-ray gas
as  tracers of  the  dark matter  potential  wells. In  this paper  we
investigate  the clustering  of dark  matter haloes  using  the second
method.

Based on  galaxy kinematics,  X-ray studies and  gravitational lensing
effects,  it is  now well  established that  clusters of  galaxies are
associated  with the  most massive  dark matter  haloes.  Observations
show  that   clusters  of   galaxies  are  strongly   clustered.   The
cluster-cluster two-point correlation function, $\xi_{\rm cc} (r)$, is
roughly  a  power  law,  $\xi_{\rm cc}(r)  =  (r_0/r)^{\alpha}$,  with
$\alpha \sim 1.8$  and with a correlation length,  $r_0$, that is much
larger than  that of  galaxies (see Bahcall  1988 for a  review; Croft
\etal 1997;  Park \& Lee 1998;  Bahcall \etal 2003).  This  is in good
agreement  with clusters  being  associated with  massive dark  matter
haloes, which are expected to be strongly clustered from the fact that
they  are associated  with high  peaks  in the  initial density  field
(e.g., Kaiser 1984).

The correlation length  of clusters is also observed  to increase with
the mean  inter-cluster separation $d  \equiv n^{-1/3}$, with  $n$ the
number density of  objects (e.g.  Bahcall \& West  1992). Since richer
clusters (where  richness expresses the number of  galaxy members) are
rarer objects, this relation between  $r_0$ and $d$ is equivalent to a
relation  between  $r_0$  and  cluster  richness.  Associating  richer
clusters with more  massive haloes, this, again, is  in good agreement
with  theoretical predictions.  Mo,  Jing \&  White (1996),  using the
halo  bias model  developed  by Mo  \&  White (1996),  found that  the
observed  $r_0$-$d$  relation  can  be  well  described  in  terms  of
halo-halo correlation  functions in  the CDM cosmogony.   In addition,
these authors showed that  the relation between correlation length and
mean  separation  can  be   used  to  constrain  models  of  structure
formation.   In particular,  they showed  that the  observed $r_0$-$d$
relation is well reproduced  by a $\Lambda$CDM model with $\Omega_{\rm
  m,0}=0.3$   and  $h=0.7$,   but  differs   significantly   from  the
predictions for  a CDM  model with $\Omega_{\rm  m, 0}=1$  and $h=0.5$
(see also Bahcall \etal 2003).

The $r_0$-$d$  relation for poorer  systems can be probed  by studying
the  correlation function  of  galaxy groups  (Zandivarez \etal  2003;
Padilla  \etal  2004). Using  groups  of  galaxies  selected from  the
2dFGRS (Eke et al. 2004), Padilla  \etal (2004) 
found that the  $s_0$-$d$ relation (with
$s_0$  the correlation length  in redshift  space) obeyed  by clusters
extends to  poor groups. Unfortunately, the  connection between galaxy
groups  and dark  matter haloes  is less  straightforward as  for rich
clusters, simply  because the smaller  number of galaxies  involved in
individual  groups  makes  it  harder  to identify  systems  that  are
physically  associated. The correspondence  between galaxy  groups and
dark  matter haloes may  therefore depend  significantly on  the group
finder  used,  complicating the  interpretation  of the  observational
results.  In order to overcome  this problem, one needs a group finder
that associates galaxies according to their common dark matter haloes.

In a recent  paper (Yang \etal 2004b, hereafter  YMBJ), we developed a
halo-based group  finder that is optimized for  grouping galaxies that
reside in  the same  dark matter halo.   We tested the  performance of
this  group  finder extensively  using  mock  galaxy redshift  surveys
constructed from the conditional  luminosity function model (Yang, Mo,
van den Bosch 2003; van den  Bosch, Yang, Mo 2003; Yang et al. 2004a),
and  found  that  our  group   finder  is  more  successful  than  the
conventional   friends-of-friends  (FOF)   algorithm   in  associating
galaxies according to their common dark matter haloes.  In particular,
our  group  finder  performs  also  reliably for  very  poor  systems,
including isolated galaxies in small mass haloes.
 
In this  paper we analyze  the 2-point correlation function  (2PCF) of
the  galaxy groups identified  by YMBJ.   As we  will show,  our group
catalogue allows  us to determine 2PCFs for  vastly different systems,
ranging from isolated galaxies  to rich clusters.  Using detailed mock
galaxy  redshift surveys  (hereafter MGRSs),  we show  that  the group
correlation  functions are  closely related  to those  of  dark matter
haloes.  The paper is  organized as follows.  In Section~\ref{sec_GC},
we briefly describe our group  finder, and summarize the properties of
the  group catalogues  obtained from  the  2dFGRS and  the MGRSs.   In
Section~\ref{sec_2pcf} we  estimate the 2PCF  of galaxy groups  in the
2dFGRS.   The relation between  the correlation  of galaxy  groups and
that     of     dark      matter     haloes     is     examined     in
Section~\ref{sec_group_halo}.    In   Section~\ref{sec_rich_corr},  we
analyze how  the correlation  length of galaxy  groups depends  on the
abundance  of  the  systems  in  consideration,  and  we  compare  the
observational     results    with    theoretical     predictions    in
Section~\ref{sec_theory}.   In  Section~\ref{sec_galaxy},  we use  our
results  to   discuss  how   one  can  understand   the  galaxy-galaxy
correlation function in terms  of the group-group correlation function
and    the    galaxy    occupation    in    groups.     Finally,    in
Section~\ref{sec_conclusion}, we summarize our results.
 
Unless  stated otherwise,  we consider  a flat  $\Lambda$CDM cosmology
with  $\Omega_m=0.3$,  $\Omega_{\Lambda}=0.7$ and  $h=H_0/(100\kmsmpc)
=0.7$   and  with   initial  density   fluctuations  described   by  a
scale-invariant power spectrum with normalization $\sigma_8=0.9$.  All
distances are calculated using this cosmology.

\section{Galaxy Groups and dark matter haloes}
\label{sec_GC}
\begin{figure*}
\centerline{\psfig{figure=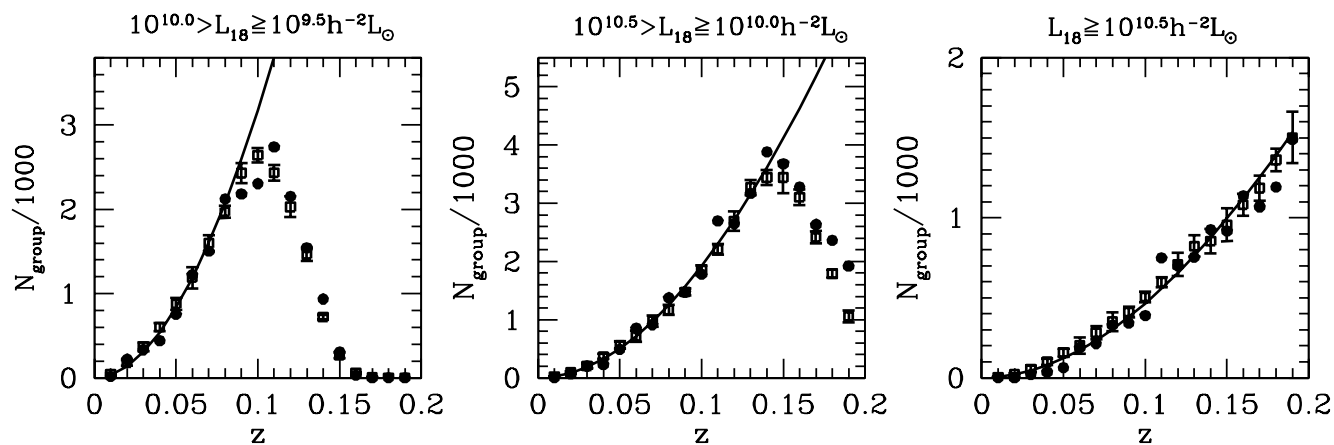,width=\hdsize}}
\caption{The redshift distributions of galaxy groups for three
  different  bins  in  $L_{18}$  (as indicated).   Open  squares  with
  errorbars are the mean and  1-$\sigma$ variance of the number counts
  for groups  in 8 independent  MGRSs, while solid dots  correspond to
  the number counts of groups in the 2dFGRS.  Solid lines indicate the
  expected value  for a  constant group number  density.  As  shown in
  YMBJ, groups  with $L_{18} \ge 10^{10.5} h^{-2}  \Lsun$ are complete
  for  $z <  0.2$. }
\label{fig:z_dist}
\end{figure*}

\subsection{The Group Finder}

In a recent study (YMBJ),  we developed a halo-based group finder that
can successfully assign galaxies into groups according to their common
haloes.  The basic idea behind our  group finder is similar to that of
the matched  filter algorithm developed  by Postman \etal  (1996) (see
also  Kepner \etal  1999;  White  \& Kochanek  2002;  Kim \etal  2002;
Kochanek \etal  2003; van den  Bosch \etal 2004a,b), although  we also
made use of the galaxy kinematics.  In summary (see YMBJ for details),
the group finder starts with  an assumed mass-to-light ratio to assign
a  tentative mass  to  each potential  group.   This mass  is used  to
estimate the size and velocity  dispersion of the underlying halo that
hosts the group,  which in turn is used  to determine group membership
(in  redshift space).   This procedure  is iterated  until  no further
changes occur in group memberships (see Appendix for more details).
   We tested the performance of our
group   finder  in  terms   of  completeness   of  true   members  and
contamination  by  interlopers,  using  detailed MGRSs.   The  average
completeness of individual  groups is $\sim 90$ percent  and with only
$\sim  20$  percent  interlopers.  Furthermore,  the  resulting  group
catalogue   is  insensitive   to   the  initial   assumption  of   the
mass-to-light ratios, and the group finder is more successful than the
conventional  FOF method (Eke et al. 2004) in associating  galaxies 
according  to their common dark matter haloes.

\subsection{2dF Groups and Mock Catalogues}

In  YMBJ we  applied the  group finder  described above  to  the final
public data release of the 2dFGRS.  This observational sample contains
about  $250,000$  galaxies  with  redshifts  and  is  complete  to  an
extinction-corrected apparent magnitude of $b_J\approx 19.45$ (Colless
et  al.  2001).   The  survey volume  of  the 2dFGRS  consists of  two
separate declination strips  in the North Galactic Pole  (NGP) and the
South Galactic  Pole (SGP),  respectively, together with  100 2-degree
fields  spread randomly  in  the southern  Galactic hemisphere.   When
identifying galaxy  groups, we  restricted ourselves only  to galaxies
with redshifts $0.01\leq z \leq 0.20$ in the NGP and SGP regions. Only
galaxies with redshift quality parameter  $q \geq 3$ and with redshift
completeness $>0.8$ were  used.  This left a grand  total of $151,820$
galaxies with  a sky  coverage of $1124  ~{\rm deg}^2$. We  obtained a
group catalogue  of $78,708$ systems which in  total contain $104,912$
galaxies.  Among these systems,  7251 are binaries, 2343 are triplets,
and 2502 are systems with four  or more members.  The vast majority of
the groups ($66,612$ systems) in  our catalogue, however,  consist of
only a single  member. Note that some faint  galaxies are not assigned
to any systems, because it is difficult to decide whether they are the
satellite galaxies of larger systems,  or whether they are the central
galaxies of small haloes.

As discussed in YMBJ, it is not reliable to estimate the (total) group
luminosity based on the assumption that the galaxy luminosity function
in groups  is similar to that  of field galaxies. We  therefore used a
more  empirical approach  to estimate  the group  luminosity $L_{18}$,
defined as  the total  luminosity of all  group members  brighter than
$M_{b_J}-5\log h = -18$. In the Appendix we describe in detail how 
$L_{18}$ is estimated for each group. As  demonstrated in detail in 
YMBJ, $L_{18}$ is tightly  correlated with the mass  of the dark  matter 
halo hosting the group,  and can be  used to rank  galaxy groups 
according  to halo masses. 
Fig~\ref{fig:z_dist}  plots  the redshift  distributions  of
groups  detected in the  2dFGRS (solid  dots) and  in our  MGRSs (open
circles). The solid line corresponds to a constant number density, and
is shown for comparison. As already shown in YMBJ, the group catalogue
is virtually complete over the entire redshift range ($0.01\leq z \leq
0.20$)  for   groups  with  $L_{18}  >   10^{10.5}  h^{-2}  L_{\odot}$
(right-hand panel).   For groups with smaller  $L_{18}$, the catalogue
is incomplete: groups with $10^{10}  h^{-2} \Lsun < L_{18} < 10^{10.5}
h^{-2} \Lsun$ are only complete to $z \sim 0.13$ (middle panel), while
those with $10^{9.5} h^{-2} \Lsun < L_{18} < 10^{10} h^{-2} \Lsun$ are
complete  down to  $z \sim  0.08$  (left-hand panel).   Note that  the
redshift  distributions of  the  MGRS agree  nicely  with the  2dFGRS,
indicating   that  we   have  properly   accounted  for   the  various
incompleteness effects  when constructing  our mock surveys  (see Yang
\etal 2004a and van den Bosch 2004a for details).
With  these  considerations,  we  can construct  volume-limited  group
samples  by  ranking all  groups  according  to  their $L_{18}$.   The
brightest $N$ groups then form a volume-limited subsample.  Using this
ranking-technique  we   construct  9  subsamples   ${\rm  O}i$,  where
$i=1,2,\cdot\cdot\cdot,9$ correspond  to different choices  of $N$ and
the  maximum  redshift  $z_{\rm  max}$.   Rather  than  characterizing
different subsamples by  $N$ and $z_{\rm max}$, we  use the mean group
separation, $d=n^{-1/3}$, where $n$ is the number density of groups in
the subsample.  Table~1 lists  the subsamples thus selected, and which
form the observational data base for our analyses.

In YMBJ, we  also applied our group finder  to eight MGRSs constructed
using  exactly the  same selection  criteria as  the 2dFGRS
(see Yang \etal  2004a for  details). Note that in the present 
MGRSs, we also include fiber collisions in a way as described in 
van den Bosch et al. (2004b). Here we  make  use of  these mock  group
catalogues to test the relation between the groups and the dark matter
haloes.  For this purpose, we generate eight dark halo catalogues from
the eight MGRSs. In which,  we  select {\it all} dark matter haloes in 
our `virtual universe' with $0.01 < z < 0.20$ that are within the area 
of the sky covered by 2dFGRS where the completeness is larger than 0.8.
Note  that  these haloes are {\it not} exactly the same as those 
corresponding to all selected groups, because the later are not 
complete due to the survey selection effect.  Subsamples of
mock groups and dark matter haloes  are constructed in the same way as
the 2dF  group samples,  i.e.  according to  the $L_{18}$  ranking for
mock groups, or according to halo mass ranking for dark matter haloes.
We denote these  two sets of subsamples as M$i$  (for mock groups) and
H$i$ (for  dark matter  haloes).  Subsamples O$i$,  M$i$ and  H$i$ all
have the  same number  of objects for  a given  $i$.  Since we  have 8
independent MGRSs, for each $i$  we have 8 independent mock subsamples
and 8  halo subsamples.  All errorbars  quoted below are  based on the
scatter among these subsamples.
\begin{table*}
\caption{The 2dFGRS group correlation functions}
\begin{tabular}{lccccccccc}
   \hline
Sample & $N$ & $z_{\rm max}$ & $d$ & $s_0(\gamma=1.8)$ &
$r_0(\gamma=1.8)$ & $r_0$ & $\gamma$ & $b/b(O6)$ & $\beta$ \\
   & & & Mpc/h & Mpc/h & Mpc/h & Mpc/h  \\
 (1) & (2) & (3) & (4) & (5) & (6) & (7) & (8) & (9) & (10) \\
\hline\hline
O1 & $250$   & $0.20$ & $43.77$ & $19.30\pm 1.73$ & --              & --      & --     & --     & -- \\
O2 & $500$   & $0.20$ & $34.74$ & $16.09\pm 1.42$ & $15.05\pm 0.83$ & $15.44$ & $2.38$ & $1.79\pm 0.11$ & $0.22\pm 0.27$\\
O3 & $1000$  & $0.20$ & $27.57$ & $14.84\pm 0.63$ & $12.79\pm 0.58$ & $13.53$ & $2.16$ & $1.58\pm 0.07$ & $0.17\pm 0.16$\\
O4 & $2000$  & $0.20$ & $21.88$ & $12.68\pm 0.41$ & $11.83\pm 0.48$ & $11.82$ & $1.79$ & $1.34\pm 0.04$ & $0.28\pm 0.15$\\
O5 & $4000$  & $0.20$ & $17.37$ & $10.84\pm 0.32$ & $9.62\pm 0.32$  & $9.36$  & $1.72$ & $1.19\pm 0.03$ & $0.36\pm 0.07$\\
O6 & $8000$  & $0.20$ & $13.79$ & $9.26 \pm 0.22$ & $8.11\pm 0.17$  & $8.02$  & $1.77$ & $1.00$         & $0.35\pm 0.08$\\
O7 & $16000$ & $0.20$ & $10.94$ & $8.12 \pm 0.17$ & $6.94\pm 0.23$  & $6.54$  & $1.68$ & $0.87\pm 0.02$ & $0.48\pm 0.07$\\
O8 & $16000$ & $0.13$ & $7.22$  & $6.24 \pm 0.20$ & $4.77\pm 0.12$  & $4.78$  & $1.86$ & $0.63\pm 0.03$ & $0.50\pm 0.12$\\
O9 & $8000$  & $0.08$ & $5.67$  & $4.55 \pm 0.40$ & $3.55\pm 0.23$  & $3.70$  & $1.85$ & $0.48\pm 0.05$ & $0.68\pm 0.26$\\
\hline
\end{tabular}
\medskip
                                                                               
\begin{minipage}{\hdsize}
  Column~(1) indicates  the sample ID. Columns~(2)  (number of groups)
  and~(3) (redshift  range: $0.01 \leq  z \leq z_{\rm  max}$) indicate
  the  selection  criteria.   Column~(4)  lists  the  mean  separation
  ($d=n^{-1/3}$) of the selected groups.  Columns~(5) and~(6) list the
  redshift  space and  real-space  correlation lengths,  respectively,
  obtained fitting  $\xi(s)$ with $\xi(s)=(s/s_0)^{1.8}  (5\le s\le 15
  \mpch)$ and  $w_p(r_p)$ using $\xi(r)=(r/r_0)^{1.8}  (3\le r_p\le 15
  \mpch)$, Column~(7) and~(8) indicate the $r_0$ and $\gamma$ obtained
  fitting  $w_p(r_p)$   using  $\xi(r)=(r/r_0)^{\gamma}$.   Column~(9)
  indicates the bias of groups  relative to that of the fiducial ${\rm
    O}6$ sample, and column~(10), finally, lists the $\beta$ parameter.
  All the errbars listed in this table are 1-$\sigma$ variances obtained  
  from the  scatter among  8 mock group samples.
\end{minipage}
                    
\end{table*}

\section{The two-point correlation function}
\label{sec_2pcf}
\begin{figure*}
\centerline{\psfig{figure=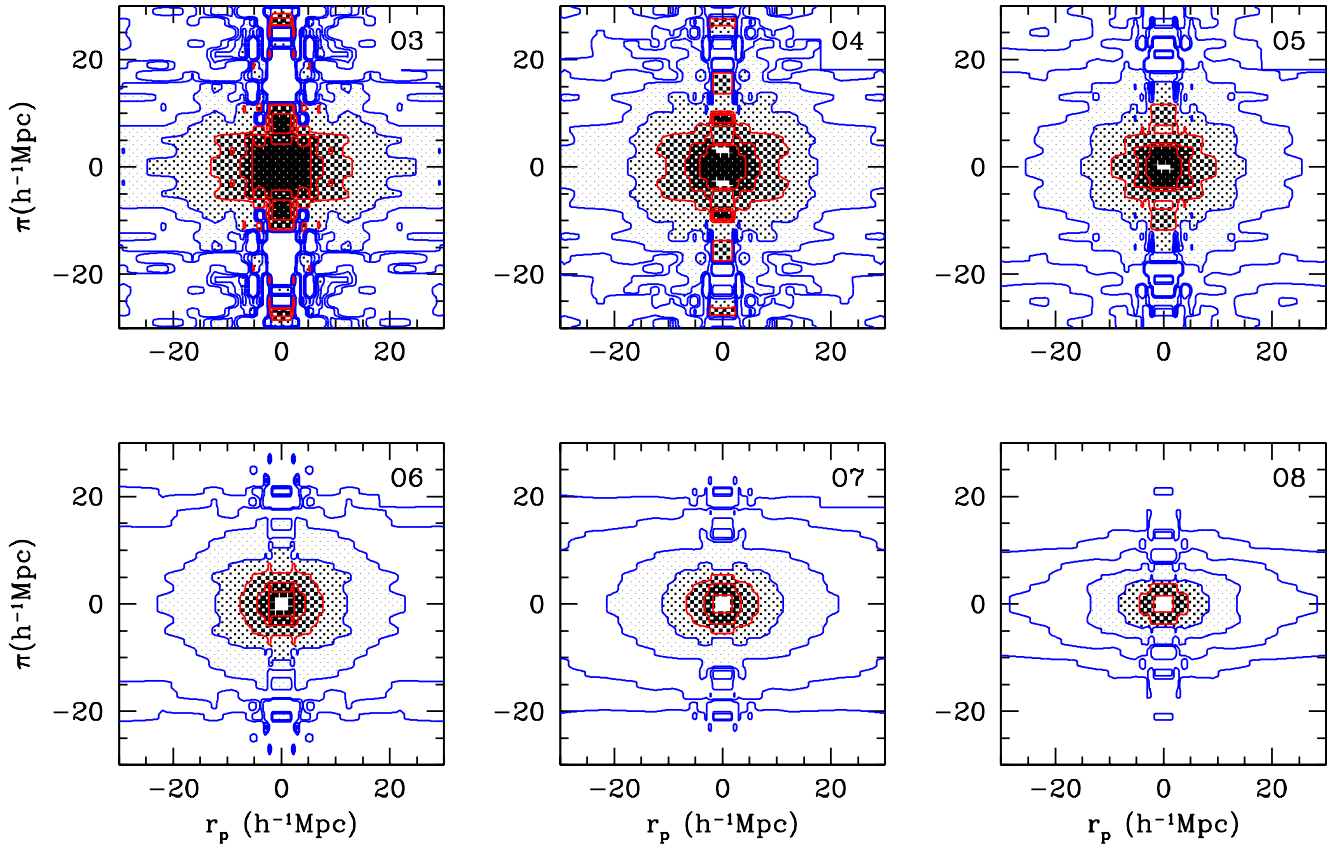,width=\hdsize}}
\caption{The two-point correlation function, $\xi(r_p,\pi)$, for 
  various  group samples  (as indicated)  extracted from  the  2dFGRS. 
  From upper left to upper right,  then lower left to lower right, the
  samples have smaller mean inter-group separations $d$ (see Table~1),
  indicating an increased inclusion of less massive systems. Note that
  samples with smaller $d$ reveal  a more pronounced flattening of the
  contours  (see also  Section~\ref{sec_rich_corr} and  the right-hand
  panel  of Fig.~\ref{fig:biasbeta}). }
\label{fig:xi2}
\end{figure*}

We compute  the group-group  (or halo-halo) 2PCF  $\xi(r_p,\pi)$ using
the following estimator
\begin{equation}
\label{tpcfest}
\xi(r_p,\pi) = {\langle RR \rangle \, \langle DD \rangle \over
\langle DR \rangle^2} - 1
\end{equation}
with  $\langle DD  \rangle$,  $\langle RR  \rangle$,  and $\langle  DR
\rangle$  the number of  group-group, random-random,  and group-random
pairs  with separation  $(r_p,\pi)$ (Hamilton  1993).  Here  $r_p$ and
$\pi$ are respectively the pair separations perpendicular and parallel
to the line-of-sight.  Explicitly, for a pair $({\bf s_1},{\bf s_2})$,
with ${\bf s_i} = c z_i {\bf \hat{r}_i}/H_0$, we define
\begin{equation}
\label{rppi}
\pi = {{\bf s} \cdot {\bf l} \over \vert {\bf l} \vert} \; ,
\;\;\;\;\;\;\;\;\;\;\;\;\;\;
r_p = \sqrt{{\bf s} \cdot {\bf s} - \pi^2}
\end{equation}
Here  ${\bf l}=(1/2)({\bf  s_1} +  {\bf s_2})$  is the  line  of sight
intersecting the pair, and ${\bf s} = {\bf s_1} - {\bf s_2}$.

Except  for ${\rm  O}1$, all  2dF  samples listed  in Table~1  contain
sufficient numbers of galaxy groups  for a proper determination of the
2PCF.    Fig~\ref{fig:xi2}    shows   the   contour-plots    for   the
$\xi(r_p,\pi)$ of  some of  these samples. Panels  from upper  left to
upper right and  from lower left to lower  right correspond to samples
${\rm O}3$ --  ${\rm O}8$. Notice that these  $\xi(r_p,\pi)$ look very
different from those of galaxies  (e.g., Hawkins \etal 2003): the only
deviation  from isotropy  is a  flattening  of the  contours at  large
separations  due to  the infall  motion induced  by  the gravitational
action of large scale structure. Unlike for galaxies, no finger-of-God
effect  on  small scales  is  present, due  to  the  fact that  groups
themselves are virialized objects rather than test particles in larger
virialized potentials.  As we will see in Section~\ref{sec_rich_corr},
this   absence  of   virial  motions   on  small   scales   makes  the
interpretation of the redshift distortion easier.
\begin{figure*}
\centerline{\psfig{figure=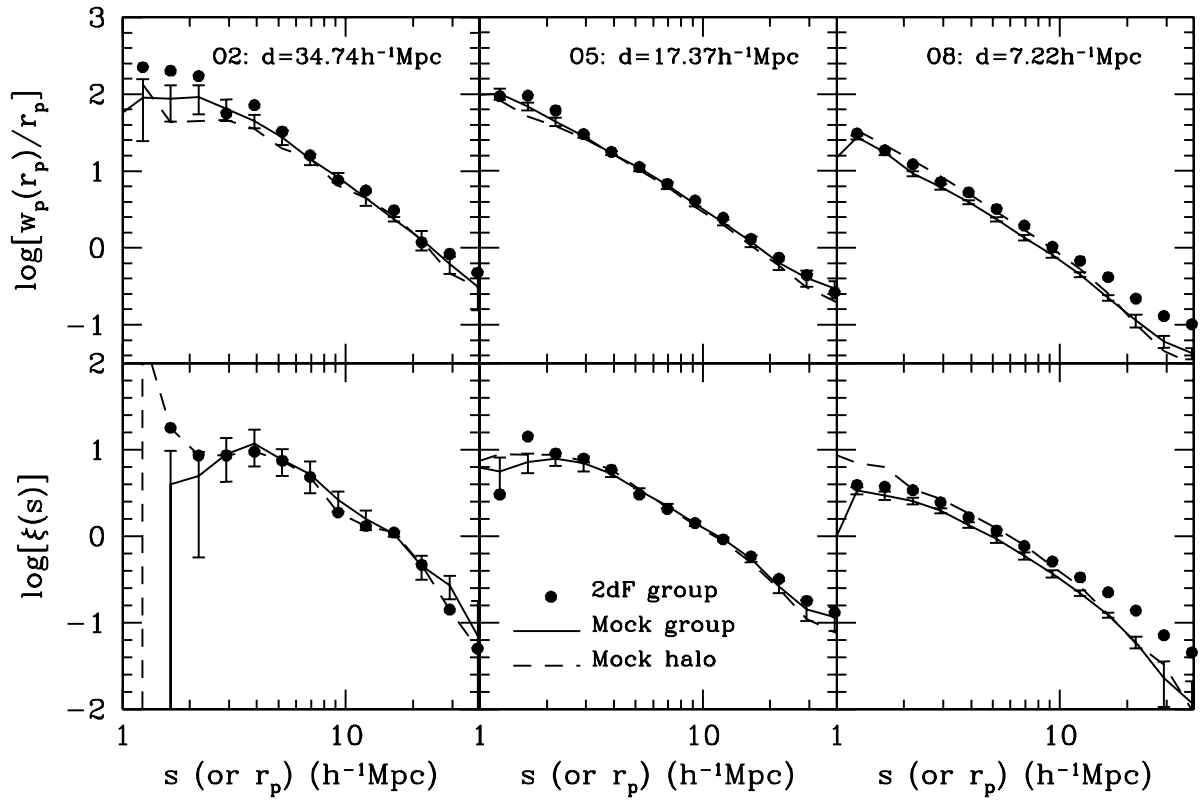,width=0.9\hdsize}}
\caption{The projected correlation function, $w_p(r_p)$ (upper
  panels),  and redshift space  correlation function,  $\xi(s)$ (lower
  panels), of groups and dark  matter haloes. Solid dots correspond to
  the groups extracted  from the 2dFGRS, while solid  and dashed lines
  indicate  the same  results but  obtained  for the  samples of  mock
  galaxies and  their corresponding dark matter  haloes, respectively. 
  The  errorbars   associated  with  the  solid   lines  indicate  the
  1-$\sigma$  variance  obtained from  the  eight  independent MGRSs.  
  Results  are shown  for three  different  values of  the mean  group
  separation, $d$, as indicated.}
\label{fig:wrp}
\end{figure*}

Since the redshift-space distortion only affects $\pi$, the projection
of  $\xi(r_p,\pi)$ along  the $\pi$  axis can  get rid  of  the infall
induced distortions and  give a function that is  more closely related
to  the   real-space  correlation  function.    This  projected  2PCF,
$w_p(r_p)$,  is related to  the real-space  2PCF, $\xi(r)$,  through a
simple Abel transform
\begin{equation}
\label{abel}
w_p(r_p) = \int_{-\infty}^{\infty} \xi(r_p,\pi) {\rm d}\pi
= 2 \int_{r_p}^{\infty} \xi(r) \,
{r \, {\rm d}r \over \sqrt{r^2 - r_p^2}}
\end{equation}
(Davis  \& Peebles  1983).  Therefore,  if  the real-space  2PCF is  a
power-law,  $\xi(r)  =  (r_0/r)^\gamma$,  the projected  2PCF  can  be
written as
\begin{equation}
\label{wxipred}
w_p(r_p) = \sqrt{\pi} \, \frac{\Gamma(\gamma/2-1/2)}{\Gamma(\gamma/2)}
\, \left( {r_0 \over r_p} \right)^{\gamma} \, r_p \,.
\end{equation}

The  black dots  in the  upper  panels of  Fig~\ref{fig:wrp} show  the
projected correlation  function $w_p(r_p)$ of  2dFGRS groups estimated
from $\xi(r_p,\pi)$ using  eq.~(\ref{abel}) with the integration range
set to $\vert \pi \vert \leq  40 \mpch$.  For comparison, we also plot
$w_p(r_p)$ for the mock groups (solid line with errorbars).  The three
panels  correspond to  samples with  mean separation  $d=34.74 \mpch$,
$17.37 \mpch$, and $7.22 \mpch$, as indicated.  Overall, the agreement
between data  and mock  is extremely good.  An exception is  the large
scales  in  sample  ${\rm  O}8$ ($d=7.22  \mpch$)\footnote{We  find  a
  similar discrepancy  between data and  model for sample  ${\rm O}9$,
  which is  limited to  an even smaller  volume, with an  even smaller
  mean separation,  than sample ${\rm O}8$,}, where  the $w_p(r_p)$ of
the mock  groups is significantly  underestimated. This is due  to the
fact that  this sample occupies a  small, nearby volume,  which in our
MGRSs  is represented  by a  small box-size  simulation that  does not
properly sample the large(r) scale structure (see Yang \etal 2004a for
details).

In  most previous  studies of  group-group correlation  functions, the
redshift-space  2PCF $\xi(s)$,  rather than  the real-space  2PCF, was
used to represent  the clustering strength (Croft \etal  1997; Park \&
Lee 1998; Zandivarez \etal 2003; Bahcall \etal 2003; 
Padilla \etal 2004), In
order to  allow for  a comparison we  also compute  the redshift-space
correlation  functions  which  are   shown  in  the  lower  panels  of
Fig.~\ref{fig:wrp}.   Here again,  the  results for  the mock  samples
match those  of the  2dFGRS samples remarkably  well, except  at large
radii in sample ${\rm O}8$.

\section{The relation between galaxy groups and dark 
matter haloes}\label{sec_group_halo}

So far we  have focused on the  2PCFs for groups in the  2dFGRS and in
our MGRSs.  We now examine  whether these results can be understood in
terms  of  2PCFs  between  dark  matter  haloes  in  the  $\Lambda$CDM
concordance cosmology.  Since the  clustering properties of CDM haloes
are well  understood (Mo  \& White 1996;  2002; Sheth \&  Tormen 1999;
Jing 1998;  Sheth, Mo  \& Tormen 2001;  Jenkins \etal 2001;  Seljak \&
Warren  2004), such  a connection  between the  populations  of galaxy
groups and dark matter haloes  enables us to understand the clustering
of groups in a cosmological context.
 
As mentioned  above, the luminosity  of a group, $L_{18}$,  is tightly
correlated to the mass of  its host halo.  Therefore, groups ranked by
the  value of $L_{18}$  may be  used to  represent dark  matter haloes
ranked  by halo  mass.   To  check this,  we  compare the  correlation
functions  of mock  group samples  (M1, M2,  etc) with  those  of dark
matter halo samples (H1, H2, etc).  The results are shown as solid and
dashed  lines,  respectively, in  Fig.~\ref{fig:wrp}.   Note that  the
correlation function of mass-ranked dark matter haloes matches that of
$L_{18}$-ranked groups remarkably well.

\begin{figure}
\centerline{\psfig{figure=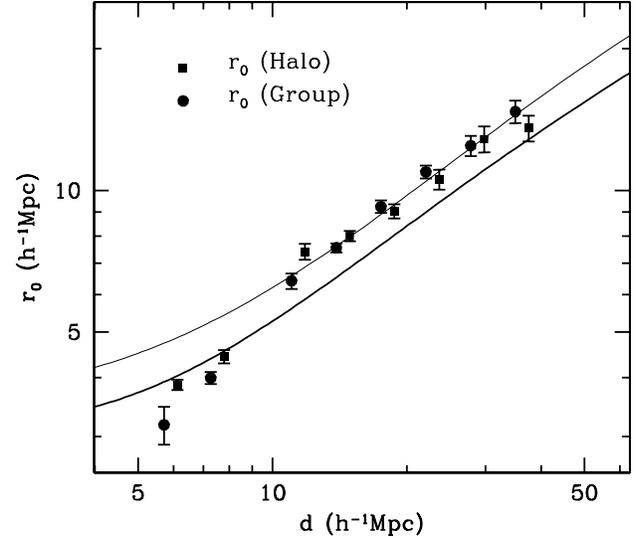,width=\hssize}}
\caption{Relation between correlation length, $r_0$, and mean
  inter-group  separation, $d$,  for groups  (solid circles)  and dark
  matter haloes  (solid squares) in the MGRSs  (errorbars indicate the
  1-$\sigma$  scatter among  the eight  independent mock  catalogues). 
  For  clarity, the  results  for  the dark  matter  haloes have  been
  shifted to  the right  by $\Delta {\rm  log} d=0.03$. Note  the good
  agreement between  groups and haloes, indicating  that groups ranked
  by luminosity can be compared  directly to dark matter haloes ranked
  by halo mass.  Thick and  thin solid lines correspond to theoretical
  predictions  based  on the  halo  bias  models  of SW04  and  SMT01,
  respectively.   Note that  the  difference between  these two  model
  predictions is  larger than the  scatter among our eight  MGRSs (see
  Section~\ref{sec_theory} for a detailed discussion).}
\label{fig:r0_mock}
\end{figure}

In  order  to  facilitate   a  more  qualitative  comparison,  we  fit
$w_p(r_p)$ with  a single  power-law of the  form~(\ref{wxipred}) over
the range $3\mpch < r_p < 15 \mpch$. The goodness of fit is based on a
simple $\chi^2$ criterion,  where the errors used for  each data point
are obtained  from the scatter  among 8 independent mock  samples (the
errors  due   to  cosmic  variance  are  typically   larger  than  the
statistical errors  on each individual  $w_p(r_p)$ measurement).  Over
the $r_p$ range  considered here, a power law is  an acceptable model. 
We treat  the slope  $\gamma$ either  as a free  parameter or  keep it
fixed at a value of $\gamma=1.8$.  In the latter case, the fit is used
to determine  only the correlation  length $r_0$.  The  differences in
the correlation lengths estimated with  fixed or free $\gamma$ is less
than  $10\%$.  Fig.~\ref{fig:r0_mock}  shows  $r_0$ (obtained  keeping
$\gamma$ fixed) as  a function of mean group  separation for both mock
groups and dark  matter haloes.  The agreement between  the groups and
dark matter  haloes is remarkably good,  especially for massive/bright
systems (note  that they have been  offset from each  other by $\Delta
\log d = 0.03$ for clarity).   At small values of $d$, i.e.  for faint
groups and low-mass haloes, the groups have slightly lower correlation
lengths than  the dark  matter haloes.  This  discrepancy is  at least
partly due to the incompleteness of the 2dFGRS (which we have mimicked
in our MGRSs).   Due to this incompleteness, which  is not present for
the  dark  matter haloes,  the  true  mean  inter-group separation  is
overestimated.   This  effect is  less  important  for larger  groups;
although some  of the member  galaxies are missed, they  still contain
sufficient members to be identified as a group.

We have also estimated  the redshift-space correlation lengths, $s_0$. 
In    this    case,   we    adopt    a    simple   power-law    model,
$\xi(s)=(s/s_0)^{1.8}$, to fit $\xi(s)$ over  the range $5 \mpch < s <
15 \mpch$. The lower limit of  $s$ adopted here is larger than that of
$r_p$   used  in  fitting   $w_p(r_p)$,  because   the  redshift-space
correlation  is  not  well  described   by  a  power  law  at  smaller
separations  (see  lower panels  of  Fig.~\ref{fig:wrp}).  As for  the
real-space  correlation  lengths,  we  find extremely  good  agreement
between the $s_0$ of mock groups and dark matter haloes (not shown).

All  these  results provide  strong  support  for  a tied  correlation
between group luminosity and halo mass, clearly demonstrating that the
groups  ranked by  luminosity  can be  compared  meaningfully to  dark
matter haloes ranked  by halo mass.
  
\section{Abundance dependence of group correlation function}
\label{sec_rich_corr}

\begin{figure*}
\centerline{\psfig{figure=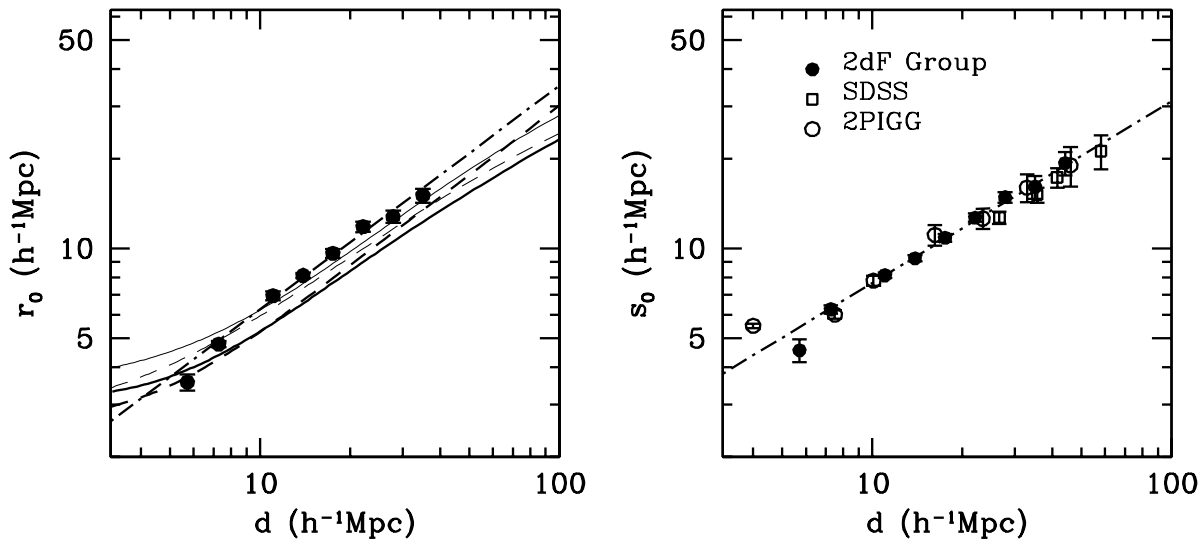,width=0.9\hdsize}}
\caption{{\it Left panel:} The relation between $r_0$ and $d$ for  
  groups selected  from the  2dFGRS (solid dots).   Errorbars indicate
  the 1-$\sigma$ variance from  8 independent mock group samples.  The
  solid  and dashed  lines are  model predictions  for  a $\Lambda$CDM
  cosmology  with $\sigma_8=0.9$ and  $0.7$, respectively.   Thick and
  thin  lines  are  based  on  the  bias models  of  SW04  and  SMT01,
  respectively.   The  dot-dashed line,  finally,  corresponds to  the
  best-fit power-law  relation, $r_0 =  1.11 \, d^{0.75}$.  {\it Right
    panel:}  The  relation   between  the  redshift-space  correlation
  length, $s_0$, and mean group  separation, $d$, for our 2dFGRS group
  catalogue (solid dots), compared to those of the SDSS (Bahcall \etal
  2003) and   2PIGG  (Padilla  \etal   2004).   The   dot-dashed  line
  corresponds to the best-fit power-law, $s_0 = 1.88 \, d^{0.61}$.}
\label{fig:r0_obs}
\end{figure*}

Having  established a  tied correlation  between group  luminosity and
halo mass, we now return to our 2dFGRS group catalogue. Using the same
fitting  procedure as  described above,  we determine  the correlation
lengths  $r_0$ and  $s_0$ as  well as  the slope  $\gamma$  for 2dFGRS
groups.   Results  are  listed   in  Table~1  (columns~5  to  8).   In
Fig.~\ref{fig:r0_obs}  we  plot the  correlation  lengths $r_0$  (left
panel) and $s_0$ (right panel) as a function of $d$, obtained assuming
a fixed  slope of $\gamma=1.8$.  For comparison,  the right-hand panel
also shows the  SDSS results (open squares) obtained  by Bahcall \etal
(2003) and the 2PIGG results  (open circles) obtained by Padilla \etal
(2004) from  the  2dFGRS.  All  three  measurements  are in  excellent
agreement  with  each  other.   Fitting our  $r_0$-$d$  and  $s_0$-$d$
relations by power laws, we obtain $r_0 = 1.11 \, d^{0.75}$ and $s_0 =
1.88 \, d^{0.61}$.  These power  laws are shown as dot-dashed lines in
Fig~\ref{fig:r0_obs}.
 
\begin{figure}
\centerline{\psfig{figure=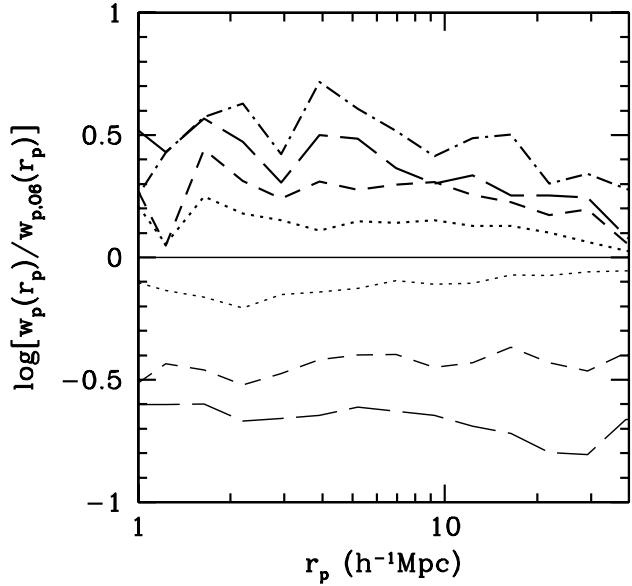,width=\hssize}}
\caption{The ratio of the projected 2PCF of various samples of 2dFGRS 
  groups relative to the fiducial sample {\rm O}6, $w_p(r_p)/w_{p,{\rm
      O}6}(r_p)$. Note  that the projected 2PCFs  of different samples
  have very similar slopes but very different amplitudes. }
\label{fig:wrp_ratio}
\end{figure}

\begin{figure*}
\centerline{\psfig{figure=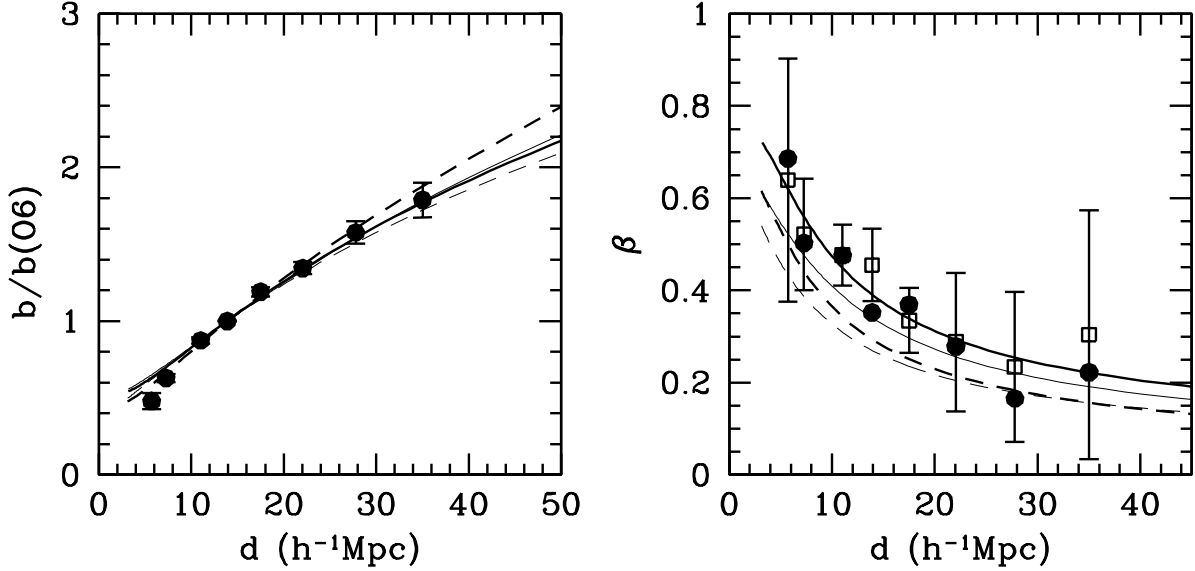,width=0.9\hdsize}}
\caption{{\it Left panel:} The relative bias, $b/b({\rm O}6)$, of
  groups  as function of  mean group  separation $d$.   These relative
  biases  are computed  from the  relation  in Fig.\ref{fig:wrp_ratio}
  using the  radial interval with  $5\mpch \le r_p \le  10\mpch$. Note
  that more massive systems (i.e.,  with larger $d$) are more strongly
  biased. {\it Right panel:} The redshift distortion parameter $\beta$
  of mock (open squares with errorbars) and 2dFGRS groups (solid dots)
  as function of $d$.  In both panels, the various lines correspond to
  the   model  predictions,   with  the   same  line   styles   as  in
  Fig~\ref{fig:r0_obs}.  Errorbars in both panels indicate
  the 1-$\sigma$ variance from  8 independent mock group samples.}
\label{fig:biasbeta}
\end{figure*}

The  correlation strength of  a sample  can also  be described  by the
ratio of  its projected  correlation function and  that of  a fiducial
sample.   Fig~\ref{fig:wrp_ratio} plots the  ratio $w_p(r_p)/w_{p,{\rm
    O}6}(r_p)$ as function of $r_p$, where ${\rm O}6$ has been used as
the  fiducial  sample.  Note  that  $w_p(r_p)/w_{p,{\rm O}6}(r_p)$  is
roughly  constant  with  $r_p$,  indicating  that  the  slope  of  the
correlation function  is roughly the  same for different  samples. The
overall {\it amplitude} of the  ratio, however, has a clear trend with
$d$. To  illustrate this we  define the {\it relative}  bias $b/b({\rm
  O}6)$ as the mean  value of the ratio $w_p(r_p)/w_{p,{\rm 0}6}(r_p)$
in the range $5\mpch \le r_p \le 10\mpch$, and plot $b/b({\rm O}6)$ as
function  of   mean  group  separation  in  the   left-hand  panel  of
Fig.~\ref{fig:biasbeta} (numerical values  are listed in column~(9) of
Table~1).

For a  given sample the value  of $s_0$ is  systematically larger than
$r_0$  (cf.  columns~5  and~6).  This  is  due to  the enhancement  of
clustering in redshift-space due  to gravitational infall. To quantify
this redshift distortion,  we use the model of  Kaiser (1987; see also
Hamilton  1992).  According  to linear  theory, the  infall velocities
around density perturbations  affect the observed correlation function
as
\begin{equation}
\label{xileg}
\xi_{\rm lin}(r_p,\pi) = \xi_0(s) {\cal P}_0(\mu) +
\xi_2(s) {\cal P}_2(\mu) + \xi_4(s) {\cal P}_4(\mu)\,,
\end{equation}
where ${\cal P}_l(\mu)$ is the $l^{th}$ Legendre polynomial, and $\mu$
is  the  cosine  of  the  angle  between  the  line-of-sight  and  the
redshift-space separation ${\bf  s}$. According to linear perturbation
theory the angular moments can be written as
\begin{equation}
\label{ang0mom}
\xi_0(s) = \left( 1 + {2 \beta \over 3} + {\beta^2 \over 5} \right) \,
\xi(r) \,,
\end{equation}
\begin{equation}
\label{ang2mom}
\xi_2(s) = \left( {4 \beta \over 3} + {4 \beta^2 \over 7} \right) \,
\left[ \xi(r) - \overline{\xi}(r) \right] \,,
\end{equation}
\begin{equation}
\label{ang4mom}
\xi_4(s) = {8 \beta^2 \over 35} \, \left[ \xi(r) +
{5 \over 2} \overline{\xi}(r) - {7 \over 2} \hat{\xi}(r) \right]  \,,
\end{equation}
with
\begin{equation}
\label{xibarr}
\overline{\xi}(r) = {3 \over r^3} \int_{0}^{r} \xi(r') r'^2 {\rm d}r'\,,
\end{equation}
and
\begin{equation}
\label{xihatr}
\hat{\xi}(r) = {5 \over r^5} \int_{0}^{r} \xi(r') r'^4 {\rm d}r'\,.
\end{equation}
In the above expressions,  $\beta$ is the linear distortion parameter,
which can  be written as $\beta=\Omega_{\rm m}^{0.6}/b$,  where $b$ is
the  bias parameter  of the  objects under  consideration.   Given the
real-space   correlation  function $\xi(r)=(r/r_0)^\gamma$,  which can   
be obtained from $w_p(r_p)$, eq.~(\ref{xileg})  can be used to  model 
$\xi(r_p,\pi)$ on linear scales. By comparing the model  predictions  
with  the  observed $\xi(r_p,\pi)$, one can easily obtain the value of 
$\beta$. We use a simple $\chi^2$ fit of the observed $\xi(r_p,\pi)$  
in the range  $8\mpch\le s \le  20\mpch$ to eq.~(\ref{xileg}) to 
probe the only  free parameter $\beta$. In the fitting, each data 
point for $\xi(r_p,\pi)$ is weighted by the error based on the 
scatter among 8 independent mock samples. The
right-hand panel  of Fig.~\ref{fig:biasbeta} plots  the $\beta$ values
thus obtained for both mock groups (squares with errorbars) and 2dFGRS
groups (solid dots).  Although  there is significant scatter, there is
a clear  trend that groups with  a smaller mean  separation $d$ (i.e.,
less  luminous  groups)  have  larger  $\beta$  and  thus  a  stronger
distorted redshift-space  correlation function (this  is also directly
visible from Fig.~\ref{fig:xi2}).  The numerical values of $\beta$ for
the 2dFGRS groups are listed  in column~(10) of Table~1, together with
the 1-$\sigma$ variances obtained from the scatter among the 8 MGRSs.

\section{Comparison with Theoretical Predictions}
\label{sec_theory}

The  tests   described  in  the   previous  sections  show   that  the
abundance-dependence  of the group-group  correlation function  can be
explained in  terms of the halo-halo correlation  function (after all,
we  constructed our group  finder to  associate galaxies  according to
their common  dark matter  halo).  This suggests that we  may 
compare  the  2dFGRS   group-group  2PCF  with  halo-halo  correlation
functions predicted by current  models of structure formation in order
to constrain cosmological parameters.

The mean number density of dark  matter haloes with mass $M>M_1$ can be
estimated through,
\begin{equation}
\label{averbias}
\overline{n}(M>M_1) = 
\int_{M_1}^{\infty} n(M) \, {\rm d} M\,,
\end{equation}
where $n(M)$ is the mass function  of dark matter haloes, which can be
estimated analytically from  the Press-Schechter formalism (e.g. Press
\& Schechter  1974; Sheth, Mo  \& Tormen 2001, hereafter  SMT01).  The
mean bias for haloes with mass exceeding $M_1$ can be estimated from
\begin{equation}
\label{averbias}
\overline{b}(M>M_1) = {1 \over \overline {n}(M>M_1)}
\int_{M_1}^{\infty} n(M) \, b(M) \, {\rm d} M\,,
\end{equation}
where $b(M)$ is the bias parameter  of dark matter haloes (Mo \& White
1996; Jing 1998;  Sheth \& Tormen 1999; SMT01;  Seljak \& Warren 2004,
hereafter SW04).  Throughout  we use the halo mass  function of SMT01,
which  has been  shown to  be  in excellent  agreement with  numerical
simulations (e.g., Jenkins \etal 2001;  White 2002). For the halo bias
parameter, we use the models of both SMT01 and SW04 for comparison.

Using $d  = n^{-1/3}$  and $r_0  = b^{2/1.8} \, r_{0,{\rm DM}}$, with
$r_{0,{\rm DM}}$  the correlation length  of the dark matter,  
we   compute   $r_0(d)$. Here the linear power 
spectrum is computed using the transfer function of Eisenstein \& Hu 
(1998), which properly accounts for the baryons, while the  non-linear 
power spectrum, which is required in calculating the dark matter 
correlation function and $r_{0,{\rm DM}}$, is computed using the 
fitting formula of Smith \etal (2003). The   solid  lines   in
Fig~\ref{fig:r0_mock} show  the model predictions  thus obtained using
the bias  models of SW04 (thick  line) and of SMT01  (thin line).  The
difference between the two bias models is quite large, and much larger
than the  1-$\sigma$ variance among  our 8 MGRSs.  Thus,  although the
$r_0$-$d$ relation can now be accurately determined from observational
data the current  models for halo bias are not  yet accurate enough to
allow  one to obtain  stringent constraints  on model  parameters (see
below).

In  the left  panel  of Fig~\ref{fig:r0_obs}  we  compare the  results
obtained for  the 2dFGRS groups with various  theoretical predictions. 
The thin solid line corresponds  to a standard $\Lambda$CDM model with
$\sigma_8=0.9$,  obtained using  the  bias model  of  SMT01. The  thin
dashed line indicates the prediction  for the same bias model but with
$\sigma_8=0.7$.   Based   on  this,   one  might  conclude   that  the
observational  data  is  in  better  agreement  with  $\sigma_8=0.9$.  
However, if  we use  the bias  model of SW04  (thick solid  and dashed
lines), the  $\sigma_8=0.7$ cosmology  matches the data  better.  Note
that although the bias model of SW04 may be more accurate than earlier
models, the  uncertainty of the bias  parameter at the  massive end is
still 10 to  20 per cent, which  is much larger than the  error on the
observational results. Clearly, the halo bias model has to be improved
further, in  order to make full  use of the constraining  power of the
present observational results.

Fig.~\ref{fig:biasbeta} compares  the theoretical predictions  for the
{\it   relative}   bias  $b/b({\rm   O}6)$   (left   panel)  and   the
redshift-distortion  parameter $\beta=\Omega_m^{0.6}/b$  (right panel,
assuming  $\Omega_m=0.3$)  with  our  observational results  from  the
2dFGRS.   Due to  the  normalization  at a  given  $d$, the  predicted
relation  between the  relative  bias  and $d$  is  quite similar  for
different models (i.e., the relative bias is fairly insensitive to the
value of  $\sigma_8$). More importantly, all model  predictions are in
good  agreement with  the observational  data. The  observed  value of
$\beta$ as a function of $d$ is also well described by the theory, but
the errorbars  are too  big to provide  stringent constraint  on model
parameters.

\section{Understanding the Shape of the Correlation Function of Galaxies}
\label{sec_galaxy}

\begin{figure*}
\centerline{\psfig{figure=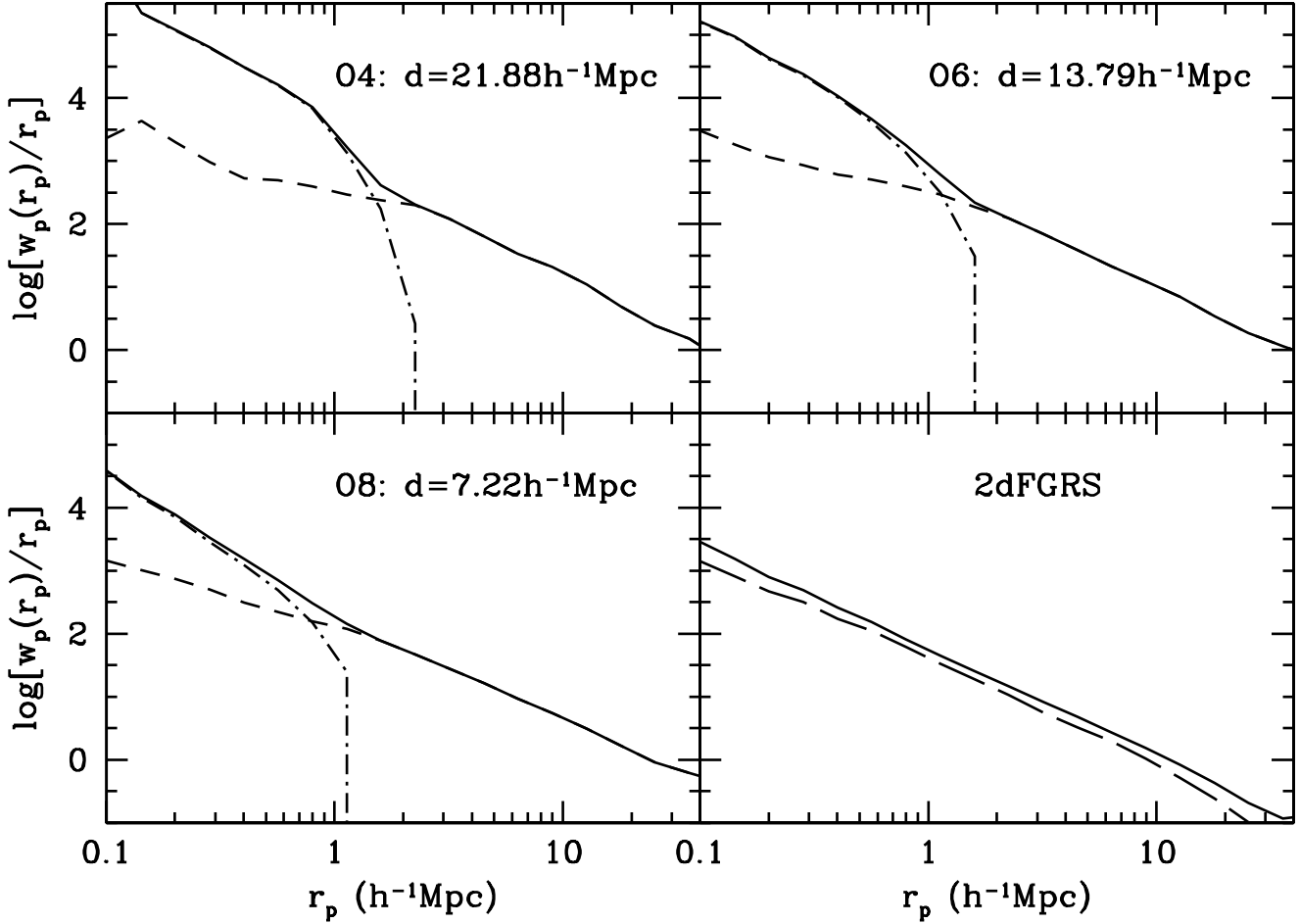,width=\hdsize}}
\caption{The projected two-point correlation function $w_p(r_p)$ of 
  galaxies in  different samples of  groups. The solid  lines indicate
  the  total  correlation functions,  $w_{p,{\rm  tot}}(r_p)$, of  all
  galaxies in group samples ${\rm O}4$ (upper left), ${\rm O}6$ (upper
  right), ${\rm  O}8$ (lower left) and  of all (151,820) galaxies  
  in the 2dFGRS sample with $0.01< z < 0.20$ and completeness $>0.8$
  (lower right). The  dot-dashed and  short dashed lines indicate  the  
  corresponding `1-halo'  terms,  $w_{p,{\rm 1h}}(r_p)$, and `2-halo' 
  terms, $w_{p,{\rm 2h}}(r_p)$, respectively.  Finally, the 
  long dashed line  in the  lower right
  panel indicates  the projected correlation function of  that half of
  all galaxies  that is not associated with the luminous groups.
   See text  for details and discussion.}
\label{fig:xig}
\end{figure*}

It is well known that  the real-space correlation function of (normal)
galaxies is remarkably well described  by a single power law for $r\la
10\mpch$.  Given  that the  {\it mass} correlation  function predicted
for typical $\Lambda$CDM cosmologies  is significantly curved on these
scales, it  is important  to understand the  origin of  this power-law
behavior.   Jing \etal  (1998) were  the first  to show  that,  if the
number of galaxies in a dark matter halo increases with halo mass as a
power  law, with  a power  index moderately  below unity,  and  if the
number  density  distribution  of   galaxies  in  massive  haloes  has
approximately  the  same profile  as  the  dark  matter, the  observed
power-law shape of the galaxy  correlation function can be reproduced. 
This kind of galaxy bias on small scales is now well understood in the
current  halo occupation model  (Peacock \&  Smith 2000;  Seljak 2000;
Scoccimarro  \etal, 2001;  Jing,  B\"orner \&  Suto  2002; Berlind  \&
Weinberg 2002;  Bullock, Wechsler  \& Somerville 2002;  Scranton 2002;
Berlind \etal,  2003; Yang, Mo \&  van den Bosch 2003;  van den Bosch,
Yang \&  Mo 2003).   In the halo  model, the  2PCF of galaxies  can be
decomposed into two terms:
\begin{equation}
\xi(r) = \xi_{\rm 1h}(r) + \xi_{\rm 2h}(r)\,,
\label{2pcf}
\end{equation}
where $\xi_{1  {\rm h}}$  represents the correlation  due to  pairs of
galaxies within the same halo  (the ``1-halo'' term), and $\xi_{2 {\rm
    h}}$  describes  the  correlation  due  to  galaxies  that  occupy
different haloes (the ``2-halo''  term).  In the standard $\Lambda$CDM
model $\xi(r)$  has a characteristic  scale at $r\sim 1$  -- $2\mpch$,
where the dominating contribution to  the 2PCF makes a transition from
the 1-halo term  to the 2-halo term. Therefore,  some departure from a
pure power law  is expected for populations of  galaxies for which the
1-halo  and  2-halo terms  are  not well  balanced.  In  fact, such  a
departure,  albeit small,  has recently  been found  in  the projected
correlation function of SDSS galaxies (Zehavi \etal 2004).

The  various analyses  in the  previous sections  have shown  that our
groups  selected from  the 2dFGRS  are nicely  related to  dark matter
haloes. Therefore,  we can directly measure the  `1-halo' and `2-halo'
terms of  the 2PCF, by simply  determining whether both  galaxies of a
pair reside in the same group (this pair adds to the 1-group term), or
whether they  reside in two different  groups (in which  case the pair
adds  to the  2-group  term). Fig.~\ref{fig:xig}  plots the  projected
two-point correlation functions of 2dFGRS galaxies that are associated
with  groups  of  different  abundances.  In  addition  to  $w_{p,{\rm
    tot}}(r_p)$, we also indicate  the `1-halo' (1-group) and `2-halo'
(2-group)  terms as  dot-dashed and  dashed lines,  respectively. Each
galaxy pair is  weighted by ${\cal W}_g =  1/(c_ic_j)$, where $c_i$ is
the survey  completeness at the  position of galaxy `$i$'.   On scales
$r_p\ga   3\mpch$  the   projected  correlation   function  $w_{p,{\rm
    tot}}(r_p)$ is  dominated by the  `2-halo' term, while  on smaller
scales ($r_p\la 1\mpch$) the  `1-halo' term dominates.  Note also that
for  galaxies  residing in  massive  systems  (i.e.,  large $d$),  the
projected correlation function clearly deviates from a pure power-law;
for these galaxies, the `1-halo' term is significantly enhanced.  This
occurs because  the `1-halo' term  contribution from a large  group is
proportional to $N_g(N_g-1)$ (with $N_g$ the number of galaxies in the
group), while the `2-halo' term contribution is proportional to $b^2$,
which  increases with halo  mass at  a slower  rate. When  adding more
galaxies  to the  sample hosted  by smaller  haloes  (i.e., decreasing
$d$),  $w_{p,{\rm  tot}}(r_p)$  becomes  better described  by  a  pure
power-law.  In  particular, as  shown by the  solid line in  the lower
right-hand panel, when all galaxies  (151,820 galaxies with $0.01< z <
0.20$ and completeness $>0.8$)  are included, the correlation function
is well represented by a single power law.

As an additional test, we estimate $w_{p,{\rm tot}}(r_p)$ for a sample
in which we remove all galaxies  in most luminous groups such that the
total number of  galaxies in the sample is  halved.  Thus, this sample
contains only  galaxies in low-mass haloes.   The correlation function
for this  sample is shown as the  long dashed line in  the lower right
panel of Fig~\ref{fig:xig}.  As for the complete sample, the projected
correlation function of this sample  is well described by a power law,
with a slightly shallower slope as for the complete sample.

Therefore,  we  conclude that  the  2PCF  of  2dF galaxies  reveals  a
power-law form as long as  sufficiently many small mass groups (halos)
are  included.  For galaxies  hosted by  massive haloes,  however, the
2PCF can deviate significantly from a pure power law.

\section{Conclusions}
\label{sec_conclusion} 
   
We  have measured  the 2PCFs  for galaxy  groups in  the  2dFGRS group
catalogue  constructed by YMBJ  using a  halo-based group  finder.  We
have  shown  that  the  current   data  allows  one  to  estimate  the
correlation function accurately for a wide range of different systems,
ranging from isolated galaxies  to rich clusters of galaxies.  Ranking
groups according to their  luminosities, $L_{18}$, we have studied how
the correlation of groups depends on group abundance.  Consistent with
previous studies  (e.g.  Bahcall \etal  2003; Padilla \etal  2004), we
found that  the amplitude of  the correlation function  increases with
group  luminosity (richness).   The dependence  of  the redshift-space
correlation length $s_0$ on the mean inter-group separation $d$ can be
quantified  as  $s_0  =   1.88  \,  d^{0.61}$,  while  the  real-space
correlation length $r_0$ reveals a somewhat steeper dependence: $r_0 =
1.11 \, d^{0.75}$.

Using  mock  group  catalogues,  obtained from  detailed  mock  galaxy
redshift  surveys, and  the  corresponding catalogues  of dark  matter
haloes, we  have shown  that the correlation  functions of  the 2dFGRS
groups  can  be understood  in  terms  of  halo-halo clustering.   The
observed correlation  length (and the corresponding bias  factor) as a
function of  group abundance is well reproduced  by associating galaxy
groups with dark matter haloes  in the standard $\Lambda$CDM model. In
particular, the  groups ranked by  $L_{18}$ match extremely  well with
dark matter  haloes ranked  by mass. We  found, however,  that current
theoretical  predictions for the  halo-halo correlation  functions are
not yet accurate  enough to allow us to  use the observational results
to put  stringent constraints on model parameters  in the $\Lambda$CDM
cosmogony.

Analyzing  the  correlation  function  for  galaxies  associated  with
different groups, we were able to bisect the 2PCF of galaxies in terms
of a  group-group correlation function and  a term due  to galaxies in
the same  group. Since our groups  are closely related  to dark matter
haloes, this  split correponds to the  1-halo and 2-halo  terms of the
correlation  function. We  have shown  how the  power-law form  of the
(projected)  correlation  function  is  broken when  only  considering
galaxies in  massive halos, and  how the balance between  the `1-halo'
and `2-halo' terms changes with halo mass.

                                                                               
\section*{Acknowledgement}

Numerical  simulations used  in this  paper  were carried  out at  the
Astronomical Data Analysis Center  (ADAC) of the National Astronomical
Observatory, Japan. We thank the 2dF team for making their data publicly
available.

\appendix
 
\section{The halo-based group finder and the assignment of $L_{18}$}

\subsection{The group finder}

In a  recent paper,  Yang \etal (2004b)  developed a  halo-based group
finder that can successfully  assign galaxies into groups according to
their common haloes. For  completeness, we present a brief description
of the group  finder here, but refer the reader  to Yang \etal (2004b)
for details.

The halo-based group finder consists of the following main steps:

{\bf  Step 1:}  Two different  methods  are combined  to identify  the
centres (and members) of potential groups. First we use the traditional
Friends-Of-Friends (FOF) algorithm with  very small linking lengths to
assign  galaxies into  groups.   The geometrical  centres  of all  FOF
groups thus  identified with  more than 2  galaxies are  considered as
centres of  potential groups. Next,  from all galaxies not  yet linked
together by  these FOF groups,  we select bright,  relatively isolated
galaxies which we also associate with the centres (and members) of 
potential groups.

{\bf Step 2:} We estimate the luminosity of a selected potential group
using
\begin{equation}\label{eq:L_grp}
L_{\rm group} = \sum_i \frac{L_i}{f_c(L_i)} 
\end{equation}
where $L_i$ is  the luminosity of each galaxy in  the group, and $f_c$
is  the incompleteness  of the  survey.  The  total luminosity  of the
group is approximated by
\begin{equation}\label{eq:L_total}
L_{\rm total} = L_{\rm group} \frac{\int_0^{\infty} L\phi(L)dL}
{\int_{L_{\rm lim}}^{\infty} L\phi(L)dL}\,,
\end{equation}
where $L_{\rm lim}$ is the minimum  luminosity of a galaxy that can be
observed at  the redshift  of the group,  and $\phi(L)$ is  the galaxy
luminosity function.

{\bf  Step  3:}  From  $L_{\rm  total}$  and a  model  for  the  group
mass-to-light  ratio,  we  compute   an  estimate  of  the  halo  mass
associated with the group in consideration. From this estimate we also
compute  the halo  radius  $r_{\rm 180}$,  the  virial radius  $r_{\rm
  vir}$,  and  the  virial velocity  $V_{\rm  vir}  =  (G M  /  r_{\rm
  vir})^{1/2}$.  The line-of-sight velocity dispersion of the galaxies
within  the  dark  matter halo  is  assumed  to  be $\sigma  =  V_{\rm
  vir}/\sqrt{2}$.

{\bf Step 4:} Once we have a group centre, and a tentative estimate of
the group size, mass, and  velocity dispersion, we can assign galaxies
to this group according to the properties of the associated halos.  If
we assume  that the phase-space distribution of  galaxies follows that
of the dark matter particles,  the number density contrast of galaxies
in redshift space  around the group centre ($=$  centre of dark matter
halo) at redshift $z_{\rm group}$ can be written as
\begin{equation}
P_M(R,\Delta z) = {H_0\over c}
{\Sigma(R)\over {\bar \rho}} p(\Delta z) \,,
\end{equation}
Here $\Delta z  = z - z_{\rm group}$ and  $\Sigma(R)$ is the projected
surface  density  of  a  (spherical)  NFW  halo,  while  the  function
$p(\Delta z){\rm  d}\Delta z $ describes the  redshift distribution of
galaxies within  the halo. See  Yang \etal (2004b) for  the functional
forms of $\Sigma(R)$ and $p(\Delta z)$ used.

Thus  defined,  $P_M(R,\Delta  z)$  is the  three-dimensional  density
contrast  in redshift  space.  In  order  to decide  whether a  galaxy
should be assigned  to a particular group we  proceed as follows.  For
each galaxy  we loop  over all groups,  and compute  the corresponding
distance $(R,\Delta z)$ between galaxy  and group centre.  Here $R$ is
the projected distance at the  redshift of the group. If $P_M(R,\Delta
z) \ge B$,  with $B=10$ an appropriately  chosen background level, the
galaxy is assigned  to the group. If a galaxy can  be assigned to more
than  one  group,  it  is   only  assigned  to  the  group  for  which
$P_M(R,\Delta z)$ has  the highest value.  Finally, if  all members of
two  groups  can be  assigned  to one  group  according  to the  above
criterion,  the  two groups  are  merged  into  a single  group.   

{\bf Step 5:}  Using the group members thus  selected we recompute the
group-centre  and go  back  to Step  2,  iterating until  there is  no
further change  in the memberships  of groups.  Note that,  unlike the
traditional FOF method, this  group finder also identifies groups with
only one member.

\subsection{The assignment of $L_{18}$ to Groups}

As discussed  in Yang et al.  (2004b), it is not  reliable to estimate
the (total) group  luminosity based on the assumption  that the galaxy
luminosity function  in groups is similar  to that of  field galaxies. 
Therefore  we used  a more  empirical approach  to estimate  the group
luminosity  $L_{18}$, defined  as the  total luminosity  of  all group
members  brighter than  $M_{b_J}-5\log  h =  -18$.  The assignment  of
$L_{18}$ to a group goes as follows.

\begin{itemize}
  
\item  We estimate  the  group  luminosity $L_{\rm  group}$  with eq.  
  (\ref{eq:L_grp})  using  only  galaxies  with $M_{b_J}-5\log  h  \le
  -18.0$.
  
\item We compute the absolute  magnitude limit $M_{b_J, lim}-5\log h $
  at the redshift of the group in consideration.
  
\item If $M_{b_J, lim}-5\log h  \ge -18.0$, then we set $L_{18}=L_{\rm
    group}$;   otherwise  $L_{18}=L_{\rm   group}  \times   f  (L_{\rm
    group},M_{b_J, lim})$,  where $f (L_{\rm  group},M_{b_J, lim})$ is
  the  correction factor  determined  from groups  at lower  redshifts
  where the  galaxy sample  is complete down  to $M_{b_J}-5\log  h \le
  -18.0$.

\end{itemize}
 
To  determine  the  correction  factor  between  $L_{\rm  group}$  and
$L_{18}$ we  first select {\it all}  groups with $z  \leq 0.09$, which
corresponds to the redshift for which a galaxy with $M_{b_J}-5\log h =
-18$ has an apparent magnitude equal to the mean limiting magnitude of
the 2dFGRS  ($b_J \leq 19.3$). By applying  further absolute magnitude
limit cuts, we estimate $L_{\rm group} (M_{b_J}\ge M_{b_J, cut})$, and
consider the  ratio $L_{18}/L_{\rm group}$  as a function  of $M_{b_J,
  cut})$ (see Fig. 9 in Yang et al. 2004b).  We fit the $L_{18}/L_{\rm
  group}$-$M_{b_J, cut}$ relation to  a functional form $[\log (L_{\rm
  group}/a_0)]^{a_1\Delta  M^2 +  a_2~\Delta  M}$, where  $\Delta M  =
M_{b_J, lim}-5\log h -18.0$, by  adjusting $a_0$, $a_1$ and $a_2$, and
use  this fit  result as  an estimate  for the  correction  factor, $f
(L_{\rm  group},M_{b_J,  lim})$.   Note  that  the  correction  factor
obtained in this way is an average, and is not expected to be accurate
for individual groups.  However, as demonstrated in detail  in Yang et
al. (2004b), $L_{18}$ so defined  is quite tightly correlated with the
mass of the dark matter halo  hosting the group in the mock catalogue,
and can be used to rank galaxy groups according to halo masses.
                                                                               

\label{lastpage}


\begin{thebibliography}{}


\bibitem[]{Bah88}
Bahcall N.A., 1988, \araa , 26, 631

\bibitem[]{Bah03}
Bahcall N.A., Dong F, Hao L., Bode P., Annis J., Gunn J.E., Schneider D.P.,
2003, \apj , 599, 814 

\bibitem[]{Bah92}
Bahcall N.A., West M.J., 1992, \apj , 392, 419

\bibitem[]{Bel02}
Berlind A.A., Weinberg D.H., 2002, \apj , 575, 587

\bibitem[]{Bel02}
Berlind A.A., \etal, 2003; \apj , 593, 1

\bibitem[]{Bul02}
Bullock J.S., Wechsler, R.H., Somerville R.S., 2002, \mnras , 329, 246

\bibitem[]{Col01}
Colless M.,  et al., 2001, \mnras , 328, 1039

\bibitem[]{Croft97}
Croft R.A.C., \etal, 1997, \mnras , 291, 305

\bibitem[]{Dav83}
Davis M., Peebles P.J.E., 1983, \apj , 267, 465

\bibitem[]{Eis98}
Eisenstein D.J., Hu W., 1998, \apj , 496, 605

\bibitem[]{Eke04}
Eke V.R., 2004, \mnras , 348, 866

\bibitem[]{Ham92}
Hamilton A.J.S., 1992, \apj , 385, L5

\bibitem[]{Ham93}
Hamilton A.J.S., 1993, \apj , 417, 19

\bibitem[]{Haw03}
Hawkins E., et al., 2003, \mnras , 346, 78

\bibitem[]{Jen01}
Jenkins A., Frenk C.S., White S.D.M., Colberg J.M. Cole S.,
Evrard A.E., Couchman H.M.P., Yoshida N., 2001, \mnras , 321, 372
                                
\bibitem[]{Jin98}
Jing Y.P., 1998, \apj , 503L, 9
                                                                           
\bibitem[]{JMB98}
Jing Y.P., Mo H.J., B\"{o}rner G., 1998, \apj , 494, 1

\bibitem[]{JBS02}
Jing Y.P., B\"{o}rner G., Suto Y., 2002, \apj , 564, 15

\bibitem[]{Kai84}
Kaiser N., 1984, \apj , 284, L9

\bibitem[]{Kai87}
Kaiser N., 1987, \mnras , 227, 1

\bibitem[]{Kepner99}
Kepner J., Fan X., Bahcall N., Gunn J., Lupton R., Xu G., 1999,
\apj, 517, 78

\bibitem[]{Koc03}
Kochanek C.S., White M., Huchra J., Macri L., Jarrett
T.H., Schneider S.E., Mader J., 2003, \apj , 585, 161

\bibitem[]{Kim02}
Kim R.J.S., et al. 2002, \aj , 123, 20

\bibitem[]{Mo96}
Mo H.J., Jing Y.P., White S.D.M, 1996, \mnras , 282, 1096 

\bibitem[]{Mo96}
Mo H.J., White S.D.M, 1996, \mnras , 282, 347

\bibitem[]{Mo02}
Mo H.J., White S.D.M, 2002, \mnras , 336, 112

\bibitem[]{Park98}
Park C., Lee S., 1998, JKAS, 31, 105

\bibitem[]{Pad04}
Padilla N.D., \etal, 2004, preprint (astro-ph/0402577)

\bibitem[]{Pea00}
Peacock J.A., Smith R.E., 2000, \mnras , 318, 1144

\bibitem[1996]{Po96}
Postman M., et al., 1996, \aj , 111, 615

\bibitem[]{PS76}
Press W.H., Schechter P., 1974, \apj , 187, 425

\bibitem[]{Sco01}
Scoccimarro R., Sheth R.K., Hui L., Jain B., 2001, \apj , 546, 20

\bibitem[]{Sc02a}
Scranton R., 2002a, \mnras , 332, 697

\bibitem[]{Sel00}
Seljak U.,2000, \mnras , 318, 203

\bibitem[]{Sel04}
Seljak U., Warren M.S., 2004, preprint (astro-ph/0403698)

\bibitem[]{SMT01}
Sheth R.K., Mo H.J., Tormen G., 2001, \mnras , 323, 1                     

\bibitem[]{She02}
Sheth R.K., Tormen G., 1999, \mnras , 308, 119

\bibitem[]{Smith03}
Smith R.E., \etal , 2003, \mnras , 341, 1311

\bibitem[]{BYM03}
van den Bosch F.C., Yang X., Mo H.J., 2003, \mnras , 340, 771 

\bibitem[]{BNMYJ}
van den Bosch F.C., Norberg P., Mo H.J., Yang X., 2004a,
preprint (astro-ph/0404033)

\bibitem[]{BNMYJ}
van den Bosch F.C., Yang X.,  Mo H.J., Norberg P., 2004b,
preprint (astro-ph/0406246)

\bibitem[2001]{WhiKoc}
White M., Kochanek C.S., 2002, \apj, 574, 24

\bibitem[]{Whi02} 
White M., 2002, \apjs , 143, 241

\bibitem[]{YMB03}  
Yang X., Mo  H.J., van  den  Bosch F.C., 2003, \mnras , 339, 1057 

\bibitem[]{Y04a}
Yang X., Mo H.J., Jing Y.P., van den Bosch F.C., Chu Y.Q., 2004a,
\mnras , 350, 1153

\bibitem[]{Y04b}  
Yang X., Mo H.J., van  den  Bosch F.C., Jing Y.P., 2004b, in press,
preprint (astro-ph/0405234) 

\bibitem[]{Zand03}
Zandivarez A., Merchan M.E., Padilla N.D., 2003, \mnras , 344, 247

\bibitem[]{Zeh04}
Zehavi I., et al., 2004, \apj , 608, 16

\end{thebibliography}
\end{document}